\documentclass[fleqn,usenatbib]{mnras}
\usepackage{newtxtext,newtxmath}
\usepackage[T1]{fontenc}
\usepackage{ae,aecompl}
\usepackage{graphicx}
\usepackage{dcolumn}
\usepackage{bm}
\usepackage{color}
\usepackage{natbib}
\usepackage{ulem}
\usepackage{booktabs}
\newcommand{\vs}{v_{\mathrm{s}}}

\newcommand{\mvir}{M_{\mathrm{vir}}}
\newcommand{\rvir}{R_{\mathrm{vir}}}

\newcommand{\rs}{r_{\mathrm{s}}}

\newcommand{\rhos}{\rho_{\mathrm{s}}}

\newcommand{\msun}{M_{\odot}}

\newcommand{\msunh}{h^{-1} M_{\odot}}
\newcommand{\mpch}{h^{-1} \mathrm{Mpc}}

\title[Distribution Functions of Dark Matter Particles]{Phase Space Distribution Functions and Energy Distributions of Dark Matter Particles in Haloes}
\author[Gross et al.]{
Axel Gross,$^{1}$\thanks{E-mail: \href{mailto:gros0408@umn.edu}{gros0408@umn.edu}}
Zhaozhou Li,$^{2}$
and Yong-Zhong Qian$^{1}$\\
$^{1}$School of Physics and Astronomy, University of Minnesota, Minneapolis, MN 55455, USA\\
$^{2}$Centre for Astrophysics and Planetary Science, Racah Institute of Physics, 
      The Hebrew University, Jerusalem, 91904, Israel}

\date{Accepted 2024 March 22. Received 2024 March 22; in original form 2024 February 05}

\pubyear{2024}
\begin{document}
\label{firstpage}
\pagerange{\pageref{firstpage}--\pageref{lastpage}}
\maketitle

\begin{abstract}
For a spherical dark matter halo with isotropic velocity distribution, the phase space distribution function (DF), the energy distribution,
and the density profile form a set of self-consistent description of its equilibrium state, and knowing one is sufficient to determine the other two.
The Navarro-Frenk-White density profile (NFW profile) is known to be a good approximation to the spherically-averaged density distribution in
simulated haloes. The DARKexp energy distribution is also known to compare well with the simulated energy distribution.
We present a quantitative assessment of the NFW and DARKexp fits to the simulated DF and energy distribution for a wide range of haloes 
in a dark-matter-only simulation from the IllustrisTNG Project. As expected, we find that the NFW fits work well except at low energy 
when the density at small radii deviates from the NFW profile. Further, the NFW and DARKexp fits have comparable accuracy 
in the region where both fit well, but the DARKexp fits are better at low energy because they require matching of the central gravitational potential.
We also find an approximate relation between the energy scale parameterizing the DARKexp energy distribution and that defined by the characteristic density
and radius of the NFW profile. This relation may be linked to the relaxation process during halo formation.
\end{abstract}


\section{Introduction}
\label{sec:intro}
We present a quantitative assessment of two types of fits to the phase space distribution function (DF) and 
the energy distribution of dark matter particles in simulated haloes.
The NFW fits are derived from the Navarro-Frenk-White density profile (hereafter NFW profile, \citealt{1997ApJ...490..493N}), 
while the DARKexp fits are based on the energy distribution proposed by \cite{2010ApJ...722..851H}.

It has been known for decades that dark matter constitutes the best explanation for a number of astronomical observations.
Notably, flattening of galactic rotation curves at large radii shows that visible galaxies must be surrounded by a large dark matter halo
(see e.g., \citealt{2019AARv.27.2S} for a review of dark matter distribution in galaxies).
As we have yet to observe dark matter directly, much of our understanding of these haloes has come from N-body simulations. 
While simulations are important, we also seek the theoretical insight and practicality that a well-constructed model can provide. 
It is well known that the NFW profile provides a good approximation to the spherically-averaged density distributions of dark matter haloes.
However, there is still a lack of consensus on how the NFW profile arises. 
In addition, this spherical density profile only represents one component of the more fundamental DF.
For a halo in equilibrium,
the DF provides a complete description because it also contains the velocity distribution of the dark matter particles. 
As the key quantity describing the dynamics of dark matter, the DF has been used for example, to discuss the annihilation of dark matter particles when the cross section depends on their energy (e.g., \citealt{Christy:2023mgs}), to study searches for dark matter particles (e.g., \citealt{2018JCAP...09..040L,2020JHEP...07..081H}), and
to infer the halo mass from satellite kinematics (e.g., \citealt{1987ApJ...320..493L,1996ApJ...457..228K,10.1046/j.1365-8711.1999.02964.x,2003A&A...397..899S,2019A&A...621A..56P,2020ApJ...894...10L}). 

Even for spherical haloes, it is very challenging to construct a DF self-consistently for a given density profile when the velocity distribution is anisotropic. 
It is therefore common to consider the simpler case of an isotropic velocity distribution, for which the DF is solely dependent on the energy and the corresponding energy distribution can be obtained using the density of energy states in a straightforward manner.
It is known that in general, the energy distribution is mostly dependent on the density profile, with weak dependence on the velocity anisotropy \citep{2021A&A...653A.140B}. Therefore, a study of the isotropic energy distribution is very informative for the more general case. 
In this paper, we focus on models of the DF and the energy distribution for haloes with a spherical density profile and an isotropic velocity distribution.

Despite many attempts (e.g., \citealt{1991MNRAS.253..414C}; \citealt{PhysRevD.73.023524}; \citealt{10.1093/mnras/stu2608}; \citealt{10.1093/mnras/stv096}), an accurate and complete form of the DF or the energy distribution for dark matter haloes remains to be found.
Given that the NFW profile is a good match to the simulated density distribution, we expect that the corresponding DF and energy distribution also match those obtained from simulations. 
This comparison represents an important check on the self-consistency of the NFW profile.
While there have been studies of the DF and the energy distribution for the NFW profile in the isotropic case \citep{2000ApJS..131...39W,2001ApJ...554.1268H} and the more general anisotropic case \citep{10.1111/j.1365-2966.2008.13441.x}, there is a lack of detailed comparisons of these results with numerical simulations. For example, \cite{Christy:2023mgs} compared the isotropic DF for a generalized NFW profile with simulations, but their focus was on the inner regions of a single simulated halo. 

In this paper, we make detailed comparisons of the DF and the energy distribution for the NFW profile with results for a wide range of haloes
in a dark-matter-only simulation from the IllustrisTNG Project \citep{10.1093/mnras/stx3304,10.1093/mnras/stx3112,10.1093/mnras/stx3040,10.1093/mnras/sty618,2018MNRAS.480.5113M}.
The NFW profile is known to fit the spherically-averaged density distributions of simulated haloes very well between radii 
$r=0.05\rvir$ and $\rvir$, where $\rvir$ is the virial radius (see \S\ref{sec:data}),
but there can be significant deviations outside this range \citep{2015MNRAS.451.1247S}. Therefore, we expect similar deviations in the DF and the energy distribution, which we seek to quantify. We find that by matching the gravitational potential at $\rvir$, the energy distribution for a simulated halo can be well described by that derived from the best-fit NFW profile. Deviations occur at low energy when the NFW profile provides a poor fit for $r<0.05\rvir$.
The comparisons of the DFs are similar to those of energy distributions, but the DF derived from the best-fit NFW profile has somewhat less accuracy 
because additional deviations are introduced through the density of energy states. We also compare the NFW fits to the simulated DFs and energy distributions with the DARKexp fits of \cite{2010ApJ...722..851H}. We find that these fits have comparable accuracy in the region where both fit well, 
and that there is an approximate relation between the energy scale of the DARKexp fits and the energy scale defined by the characteristic density and radius of the NFW profile. 
The DARKexp fits are better at low energy because they require matching of the central gravitational potential.

The rest of the paper is organized as follows. In \S\ref{sec:model}, we outline the relationship among the DF, the energy distribution, and the density profile. In \S\ref{sec:data}, we describe the sample of simulated haloes, their best-fit NFW profiles, and their energy distributions and DFs constructed from simulations.
In \S\ref{sec:result}, we discuss the NFW fits to the simulated energy distributions and DFs. In \S\ref{sec:DARKexp}, we discuss the DARKexp fits to the simulated energy distributions and compare the corresponding density profiles and DFs with simulations and the NFW fits. We summarize our results and give conclusions in \S\ref{sec:sum}.

\section{DF, Energy Distribution, and Density Profile}
\label{sec:model}
For a spherical halo with an isotropic velocity distribution, the DF is a function of the energy per unit mass
$E=v^2/2+\Phi(r)$ only, where $v$ is the velocity and $\Phi(r)$ is the gravitational potential as a function of the radius $r$. 
For convenience of presentation below, we take $\Phi(\rvir)=0$.
We consider the DF as a mass distribution in phase space:
\begin{align}
    f(E)=\frac{d^6M}{d^3\mathbf{r}d^3\mathbf{v}}=m_{\rm d}\frac{d^6N}{d^3\mathbf{r}d^3\mathbf{v}},
\end{align}
where $m_{\rm d}$ is the mass of a dark matter particle. The corresponding density profile is given by
\begin{align}
\label{eq:densityfromfe}
    \rho(r)=\int d^3\mathbf{v}\,f(E)=4\pi\int_{\Phi(r)}^{\Phi(\infty)} dE\,f(E)\sqrt{2[E-\Phi(r)]}.
\end{align}
For consistency, $\Phi(r)$ and $\rho(r)$ satisfy Poisson's equation $\nabla^2\Phi(r)=4\pi G\rho(r)$, where $G$ is the gravitational constant.
For the NFW profile given by
\begin{align}
    \rho(r)=\frac{\rhos}{(r/\rs)(1+r/\rs)^2},
\end{align}
where $\rhos$ and $\rs$ are the characteristic scales for density and radius, respectively,
the gravitational potential is
\begin{align}
    \Phi(r)=\vs^2\left[\frac{\ln(1+\rvir/\rs)}{\rvir/\rs}-\frac{\ln(1+r/\rs)}{r/\rs}\right],
\label{eq:nfwphi}
\end{align}
where $\vs=\rs\sqrt{4\pi G\rhos}$. 

Using the well-known procedure first proposed by \citet{10.1093/mnras/76.7.572},
we can invert Eq.~(\ref{eq:densityfromfe}) to obtain the DF for a given $\rho(r)$:
\begin{align}
\label{eq:eddington}
    f(E)=\frac{1}{\pi^2\sqrt{8}}\frac{d}{dE}\int_{r_E}^{\infty}
    \frac{dr}{\sqrt{\Phi(r)-E}}\frac{d{\rho}}{dr},
\end{align}  
where $r_E$ corresponds to $\Phi(r_E)=E$.

The energy distribution $dM/dE$ can be obtained from the DF as
\begin{equation}
    \frac{dM}{dE}=\int d^3\mathbf{r} d^3\mathbf{v}\,f(E)\delta \left(\frac{v^2}{2}+\Phi(r)-E\right)=f(E)g(E),
\label{eq:dmde}
\end{equation}
where
\begin{equation}
    g(E)=\int d^3\mathbf{r} d^3\mathbf{v}\,\delta \left(\frac{v^2}{2}+\Phi(r)-E\right)
\label{eq:ge}
\end{equation}
is the density of energy states. 
For simulated haloes, we will obtain the energy distribution directly by counting the particles inside the virial radius $\rvir$.
Therefore, for comparisons with the simulated energy distribution, the density of energy states should be 
\begin{align}
\label{eq:geupdated}
g(E)=16\pi^2\int_0^{r_{E}^*} dr\,r^2 \sqrt{2[E-\Phi(r)]},
\end{align}
where $r_E^*=\text{min}(r_E,\rvir)$. The generalization of $g(E)$ for other outer boundary radii is straightforward.

From the above discussion, the DF $f(E)$, the energy distribution $dM/dE$, and the density profile $\rho(r)$ are a set of self-consistent
description of a halo. In principle, knowing one is sufficient to determine the other two. Following the presentation of simulated haloes in \S\ref{sec:data},
we will obtain the energy distribution and the DF from the best-fit NFW profile in \S\ref{sec:result}. The procedure is evident from
Eqs.~(\ref{eq:nfwphi}), (\ref{eq:eddington}), (\ref{eq:dmde}), and (\ref{eq:geupdated}).

Based on tentative but intriguing arguments of equilibrium statistical mechanics, \cite{2010ApJ...722..851H} derived the DARKexp energy
distribution
\begin{equation}
\frac{dM}{dE}=A\left[\exp\left(\frac{E-\Phi_0}{\sigma^2}\right)-1\right],
\end{equation}
where $A$ is a normalization factor, $\Phi_0$ is the central gravitational potential, and $\sigma^2$ is the characteristic energy scale. 
It has been shown that the above energy distribution and the associated density profile match those obtained from simulations \citep{Williams_20101,Williams_20102,2015ApJ...811....2H,2016JCAP...09..042N}. The DARKexp density profile was also
compared with lensing observations of galaxy clusters by \cite{2016ApJ...821..116U} and explored theoretically by \cite{Destri2018}.
We will describe the more complicated procedure to obtain the density profile and the DF from the DARKexp energy distribution in \S\ref{sec:DARKexp}.

\section{Simulated Haloes}
\label{sec:data}
We use the simulated haloes from the IllustrisTNG Project \citep{10.1093/mnras/stx3304,10.1093/mnras/stx3112,10.1093/mnras/stx3040,10.1093/mnras/sty618,2018MNRAS.480.5113M}.
Specifically, we select the TNG300-1-Dark (hereafter TNG300) run for its better coverage of massive clusters. We have confirmed that
using the higher-resolution TNG100-1-Dark run gives very similar results for the smaller ($\sim10^{12}\msun$) haloes.
The TNG300 run was carried out in a periodic box of size $L=205\mpch$ with
$2500^3$  dark matter particles, each with a mass $m_{\rm d}=7\times10^7\msunh$; thus, a halo of $10^{12}\msunh$ contains $\approx 1.4\times 10^4$ particles.
The simulations were performed with the Planck Collaboration XIII \citep{2016A&A...594A..13P} cosmological parameters:
$\Omega_\mathrm{m} =  0.3089$, $\Omega_\mathrm{b} = 0.0486$, $\Omega_{\Lambda} = 0.6911$, 
$h=0.6774$, $n_\mathrm{s}=0.9667$, and $\sigma_8 = 0.8159$.

We focus on isolated haloes with virial mass $\mvir \in [10^{12},10^{14.5}]\msunh$. 
The virial mass and radius, $\mvir$ and $\rvir$, respectively, are defined such that the average density inside $\rvir$ is equal to $\Delta_\mathrm{vir}$ times the critical density of the Universe,
where $\Delta_\mathrm{vir}=18\pi^2+82(\Omega_\mathrm{m}-1)-39(\Omega_\mathrm{m}-1)^2\approx 102$ is the virial factor \citep{1998ApJ...495...80B}.
We define an isolated halo by the following criteria: its vicinity within $2.5\rvir$ does not contain any subhalo of mass exceeding $0.1\mvir$
and does not overlap with the $2.5\rvir$ vicinity of any halo of mass exceeding $0.5\mvir$. This selection is not very restrictive in that 
it removes $\sim 40\%$ of the $\sim 33,000$ haloes with $M_\mathrm{vir}>10^{12}\msunh$. 

From the above TNG300 sample, we randomly select 100 haloes and check if they are in equilibrium to good approximation.
We define the relaxation level by the offset $\Delta r=|\bm{r}_\mathrm{CoM} - \bm{r}_\mathrm{MinPot}|$ 
between the center of mass and the location of the minimum potential. Applying $\Delta r < 0.01\rvir$
removes 19 haloes. The above criterion is much more stringent than some criteria proposed by others
(e.g., $\Delta r<0.07 \rvir$, see \citealt{2007MNRAS.378...55M,10.1111/j.1365-2966.2007.12381.x}),
and the reason for its adoption will be explained in \S\ref{sec:scale}. We remove another two haloes that are
insufficiently relaxed based on the NFW fits to their density profiles (see \S\ref{sec:nfw}). The remaining 79 haloes are studied in detail.
Although we show the results for only eight representative haloes in Figs.~\ref{fig:density}, \ref{fig:dMdE}, \ref{fig:fE}, and \ref{fig:potential} below, these results should represent
the general sample of isolated and relaxed haloes with $\mvir \in [10^{12},10^{14.5}]\msunh$.

The energy for constructing the energy distribution $dM/dE$ and the DF $f(E)$ is very sensitive to 
the choice of reference frame. Guided by \cite{2012MNRAS.427.2437H}, we use the mean values of the most bound 10\% 
of the particles in a halo to define its position and velocity. This choice mostly affects the energy of particles 
in the central region but does not affect those in the outer halo significantly.
\subsection{NFW Fit to Density Profile}
\label{sec:nfw}
For each halo, we calculate the spherically-averaged density profile, and 
use the data in the radial range of $[0.05, 1] \rvir$ to obtain the parameters $\rhos$ and $\rs$ for the NFW profile \citep{1997ApJ...490..493N}, 
which is known to accurately match simulations in the above region \citep{2015MNRAS.451.1247S}.
The best-fit parameters are obtained with the Python package \textsc{lmfit} \citep{newville_matthew_2014_11813}
by minimizing the mean square difference in $\log \rho$ between the NFW fit and the data. 
The root of the minimum mean square difference corresponding to the best-fit NFW profile is denoted as $\delta_{\log \rho}$.
The uncertainties in $\rhos$ and $\rs$ are estimated through inversion of the Hessian matrix, which provides non-rigorous but usually reasonable estimates of errors.
We also calculate the concentration $c=\rvir/\rs$ and estimate its uncertainty via the propagation of errors.
As mentioned above, two haloes with $(c, \mvir,\delta_{\log \rho})=(2.9,3.2\times10^{13} \msunh, 0.26)$ and $(4.4,4.6\times10^{13} \msunh, 0.23)$
are removed because they are poorly fitted by the NFW profile. 
Their low concentration suggests a late formation time and insufficient relaxation \citep{2009ApJ...707..354Z}.
The remaining 79 haloes of our selected sample have an average $\delta_{\log \rho}$ value of 0.07 dex, which is $\approx 3.3$--3.7 times smaller than those
for the two removed haloes. 

\begin{figure*}
\centering
\includegraphics[width=1\textwidth]{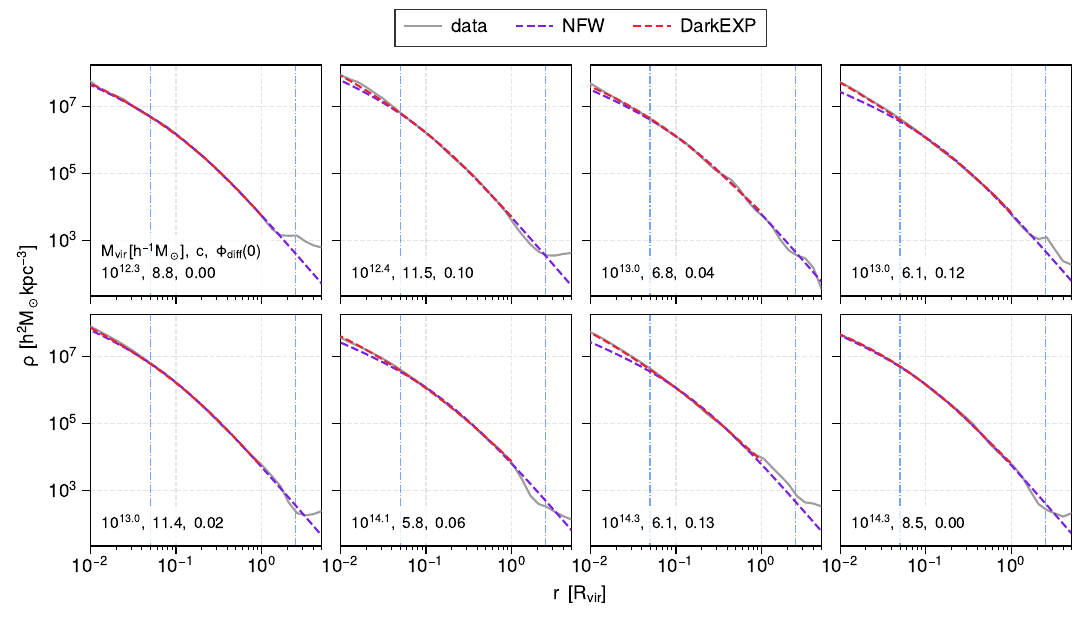}
\vspace{-2.5em}
\caption{%
Comparison of density profiles for the representative sample of haloes. In each panel, the grey curve shows the spherically-averaged density profile of a simulated halo, the purple curve shows the best-fit NFW profile,
and the red curve shows the result for the best-fit DARKexp energy distribution. The blue lines indicate $r=0.05\rvir$ and $2.5\rvir$. The parameter $\Phi_{\text{diff}}(0)$ 
is in units of $\vs^2$ and refers to the difference $\Phi_{\text{NFW}}(0)-\Phi_{\text{sim}}(0)$ between the central potential of the NFW fit and the simulated result.
}
\label{fig:density}
\end{figure*}

We compare the best-fit NFW profiles with the data for eight representative haloes in Fig.~\ref{fig:density}.
As expected, the NFW profiles match the data very well for $r\in [0.05, 1] \rvir$.
Typically, there are discrepancies at $r<0.05\rvir$. In addition, most of the haloes show gradual deviations from the NFW profile 
at $r>\rvir$ and there is a change in the slope at $r\approx 2.5\rvir$. This location can be identified as the depletion radius,
which separates a growing halo from its draining environment \citep{2021MNRAS.503.4250F}.
While the exact location of the depletion radius depends on the formation history of the halo, 
its typical value is $\approx 2.5\rvir$ \citep{2021MNRAS.503.4250F,2023ApJ...953...37G}.

\subsection{Simulated Energy Distribution and DF}
\label{sec:simdf}
\begin{figure*}
\centering
\includegraphics[width=1\textwidth]{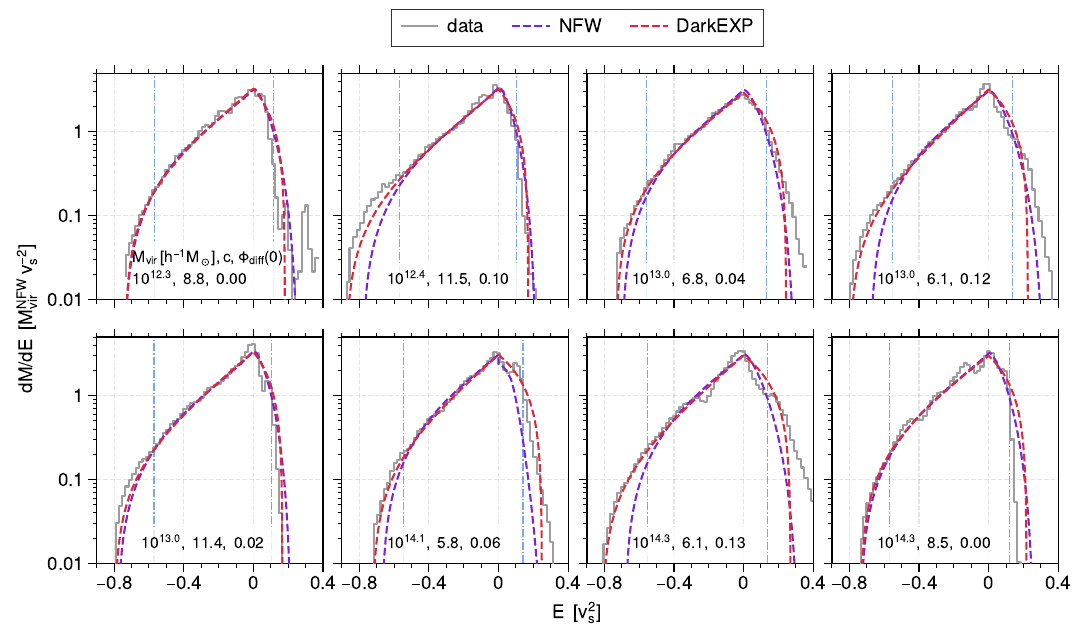}
\vspace{-2.5em}
\caption{%
Comparison of energy distributions for the representative sample of haloes. In each panel, the grey curve shows the result from simulations, the purple curve shows the result from the best-fit NFW profile, 
and the red curve shows the best-fit DARKexp energy distribution. The blue lines indicate $E=\Phi_{\text{NFW}}(0.05\rvir)$ and $\Phi_{\text{NFW}}(2.5\rvir)$.
The energy distributions fall off at high energies because only particles inside $R_{\rm vir}$ are counted. The parameter $\Phi_{\text{diff}}(0)=\Phi_{\text{NFW}}(0)-\Phi_{\text{sim}}(0)$ is in units of $\vs^2$.
}
\label{fig:dMdE}
\end{figure*}

\begin{figure*}
\centering
\includegraphics[width=1\textwidth]{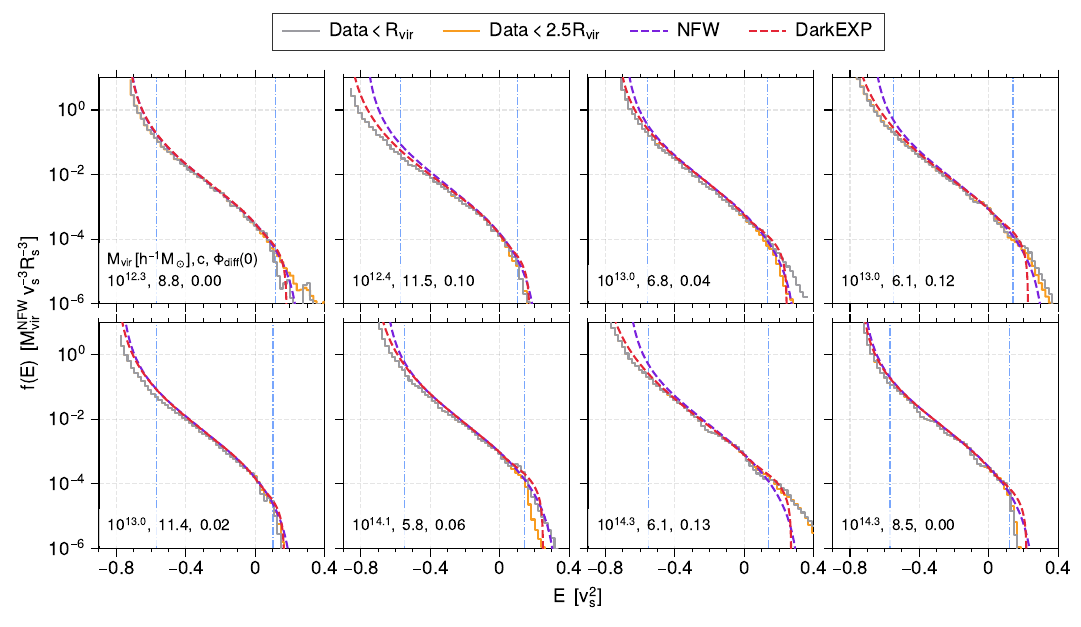}
\vspace{-2.5em}
\caption{%
Comparison of DFs for the representative sample of haloes. In each panel, the grey and orange curves show the results from simulations using particles inside $R_{\rm vir}$ and $2.5R_{\rm vir}$, respectively, the purple curve shows the result from the best-fit NFW profile, and the red curve shows the result from the best-fit DARKexp energy distribution. The blue lines indicate $E=\Phi_{\mathrm{NFW}}(0.05\rvir)$ and $\Phi_{\mathrm{NFW}}(2.5\rvir)$. The parameter $\Phi_{\text{diff}}(0)=\Phi_{\text{NFW}}(0)-\Phi_{\text{sim}}(0)$ is in units of $\vs^2$.
}
\label{fig:fE}
\end{figure*}

While the DF is more fundamental, to construct it directly from current simulations is generally impractical because the number of particles used is insufficient to sample the six-dimensional phase space in detail (cf. \citealt{2006MNRAS.373.1293S}). In contrast, current simulations have a sufficient number of particles to give an accurate description of the energy distribution $dM/dE$. Following e.g., \cite{1997MNRAS.286..329N}, we first construct $dM/dE$ directly from simulations and then obtain the DF $f(E)$ from Eq.~(\ref{eq:dmde}) using the density of energy states $g(E)$ in Eq.~(\ref{eq:geupdated}). Here we calculate $g(E)$ using the $\Phi(r)$ corresponding to the simulated $\rho(r)$.
For each halo, we construct $dM/dE$ by counting particles inside the virial radius $\rvir$ and calculating the energy of each particle.
The simulated energy distributions are shown in Fig.~\ref{fig:dMdE} for the eight representative haloes.
The corresponding DFs are shown in Fig.~\ref{fig:fE}.

To confirm that the selected haloes are in equilibrium to good approximation, we repeat the construction of $dM/dE$ and $f(E)$ by counting particles inside $2.5\rvir$. These DFs are also shown in Fig.~\ref{fig:fE} for the eight representative haloes. It can be seen that the simulated DFs for particles inside $\rvir$ and $2.5\rvir$ are consistent up to $E\approx\Phi(2.5\rvir)$, which indicates that the selected haloes are relaxed up to $2.5\rvir$.

\section{NFW Fits to Energy Distribution and DF}
\label{sec:result}
\begin{figure*}
\centering
\includegraphics[width=1\textwidth]{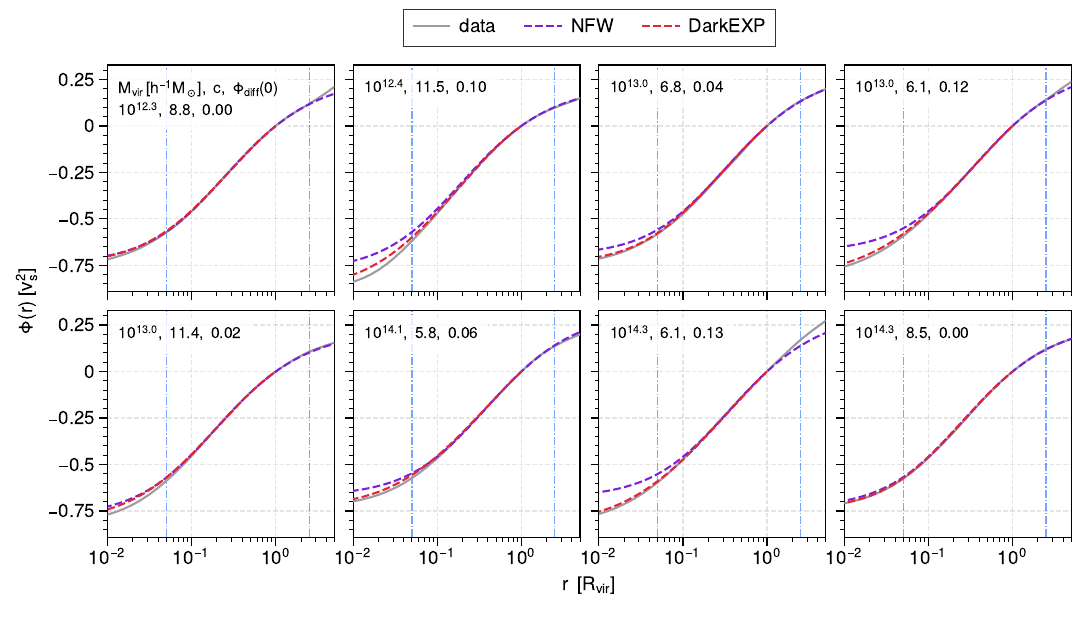}
\vspace{-2.5em}
\caption{%
Comparison of the potentials for the representative sample of haloes. In each panel, the grey curve shows the result from simulations, the purple curve shows the result from the best-fit NFW profile, and the red curve shows the result from the best-fit DARKexp energy distribution. The blue lines indicate $r=0.05\rvir$ and $2.5\rvir$. The parameter $\Phi_{\text{diff}}(0)=\Phi_{\text{NFW}}(0)-\Phi_{\text{sim}}(0)$ is in units of $\vs^2$.
}
\label{fig:potential}
\end{figure*}

In this section, we compare the energy distribution and the DF calculated from the best-fit NFW profile for a halo with the simulated results.
Because the NFW potential in Eq.~(\ref{eq:nfwphi}) tends to deviate significantly from the simulated potential outside the radial range $[0.05\rvir,2.5\rvir]$
(see Fig.~\ref{fig:potential}), it is non-trivial to match the particle energy in the NFW potential to that in the simulated halo.
In view of the differences between the NFW and simulated potentials at both small and large radii, we choose to match the energy at $\rvir$, which is effectively accomplished by setting both potentials to zero at $\rvir$ as we have done. With this prescription, for the same $E$, particles in the NFW and simulated potentials move in approximately the same region for the majority of the energy range $[\Phi(0),\Phi(2.5\rvir)]$. For convenience of comparing fits to energy distributions and DFs with simulations, we use the energy range defined by the NFW potential $\Phi_{\rm NFW}(r)$.

\subsection{Comparison of Fitted and Simulated Energy Distributions}
\begin{figure}
\centering
\includegraphics[width=\columnwidth]{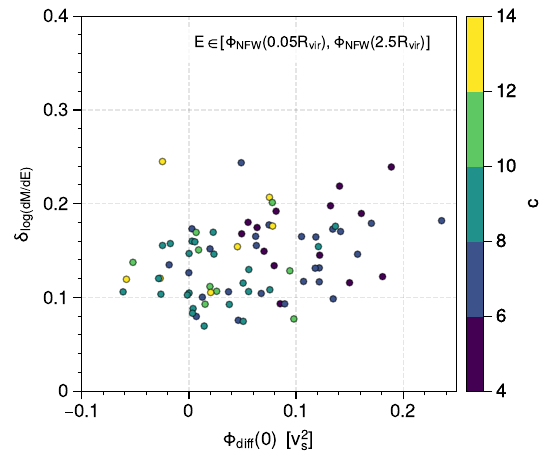}
\vspace{-2em}
\caption{%
Root mean square error $\delta_{\log(dM/dE)}$ of the best-fit NFW energy distribution as a function of $\Phi_{\text{diff}}(0)=\Phi_{\text{NFW}}(0)-\Phi_{\mathrm{sim}}(0)$ and the concentration $c$. Each symbol represents one of the 79 selected haloes.
}
\label{fig:fitstat}
\end{figure}

As mentioned in \S\ref{sec:model}, we use Eqs.~(\ref{eq:nfwphi}), (\ref{eq:eddington}), (\ref{eq:dmde}), and (\ref{eq:geupdated}) to calculate
the $dM/dE$ from the best-fit NFW profile for each halo. We only count particles inside $\rvir$. The results are compared with the simulated ones for the
eight representative haloes in Fig.~\ref{fig:dMdE}. It can be seen that the NFW fits describe the simulated energy distributions very well for the energy range
$[\Phi_{\mathrm{NFW}}(0.05\rvir),\Phi_{\mathrm{NFW}}(\rvir)]$. 
Specifically, the average root mean square error in $\log(dM/dE)$ is $\delta_{\log(dM/dE)}=0.11$~dex across our entire selected sample of 79 haloes.
The error slightly increases to $\delta_{\log(dM/dE)}=0.14$~dex for the somewhat wider energy range 
$[\Phi_{\mathrm{NFW}}(0.05\rvir),\Phi_{\mathrm{NFW}}(2.5\rvir)]$ (see Fig.~\ref{fig:fitstat}).

\begin{figure}
\centering
\includegraphics[trim={2.5mm 0 2mm 0},clip,width=\columnwidth]{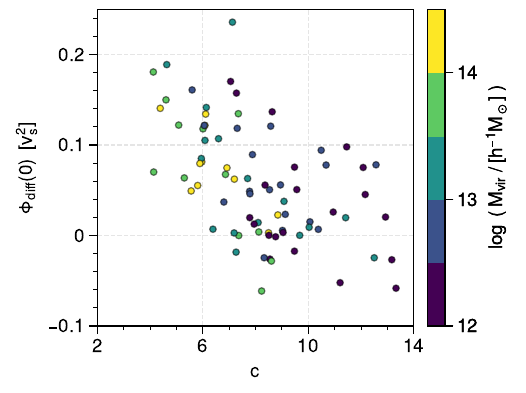}
\vspace{-2em}
\caption{%
Difference in the central potential $\Phi_{\text{diff}}(0)=\Phi_{\text{NFW}}(0)-\Phi_{\mathrm{sim}}(0)$ as a function of the concentration $c$ and the virial mass $\mvir$. Each symbol represents one of the 79 selected haloes.
}
\label{fig:psidiff}
\end{figure}

Because the NFW profile is no longer a good description of simulated haloes at $r>2.5\rvir$, the NFW fits deviate from the simulated energy distributions for $E\gtrsim\Phi_{\mathrm{NFW}}(2.5\rvir)$
(see Fig.~\ref{fig:dMdE}). Similarly, deviations occur for $E\lesssim\Phi_{\mathrm{NFW}}(0.05\rvir)$, which can be traced to differences between the NFW and simulated central potentials.
The difference $\Phi_{\rm diff}(0)=\Phi_{\rm NFW}(0)-\Phi_{\rm sim}(0)$ is shown as a function of the concentration $c$ and the virial mass $\mvir$ 
in Fig.~\ref{fig:psidiff}.
It can be seen that $\Phi_{\rm diff}(0)$ tends to be larger for haloes with higher $\mvir$ and lower $c$, which typically form later \citep{2009ApJ...707..354Z}. This result suggests that deviations from the NFW profile are largely due to insufficient time for full relaxation.
For haloes with large values of $\Phi_{\text{diff}}(0)$, Fig.~\ref{fig:dMdE} shows that the NFW fits to $dM/dE$ fall off too steeply with decreasing energy, which leads to large deviations from the simulated energy distributions for the most tightly-bound particles. This discrepancy can be understood
from the approximate relation $dM/dE\propto r_E$ (Gross, Li, Qian, in preparation). Recall that $r_E$ is the inverse function of $\Phi(r_E)=E$. The more slowly $\Phi(r)$ decreases with decreasing $r$, the more steeply $r_E$ decreases with decreasing $E$. Figure~\ref{fig:potential} shows that
the NFW potential tends to be flatter at small radii. Therefore, the NFW fits to $dM/dE$ should decrease more steeply
with decreasing energy than the simulated energy distributions.

There can be some (up to $\sim 10\%$) differences in the virial mass between simulated haloes and their best-fit NFW profiles.
For clarity, we use the NFW virial mass $\mvir^{\rm NFW}$ for the units of $dM/dE$ and $f(E)$ in Figs.~\ref{fig:dMdE} and \ref{fig:fE}, respectively.

\subsection{Comparison of Fitted and Simulated DFs}
For simulated haloes, we construct the energy distributions directly from simulations and then obtain the DFs from Eq.~(\ref{eq:dmde}) using the density of energy states (see \S\ref{sec:simdf}).
This procedure depends on the outer boundary radius inside which particles are counted for the energy distribution.
In contrast, for the NFW profile, we obtain the DF directly from Eq.~(\ref{eq:eddington}) without considering any outer boundary radius for counting particles.
We compare the NFW fits with the simulated DFs for the eight representative haloes in Fig. \ref{fig:fE}. 
It can be seen that the NFW fits describe the simulated DFs relatively well for the energy range
$[\Phi_\mathrm{NFW}(0.05\rvir),\Phi_\mathrm{NFW}(2.5(\rvir)]$. Specifically, the average root mean square error in $\log f(E)$ is $\delta_{\log f(E)}=0.24$~dex across our entire selected sample of 79 haloes.
This average error is larger than that (0.14~dex) for the NFW fits to the energy distributions, which can be understood from the relation $f(E)=(dM/dE)/g(E)$.
Because the density of energy states $g(E)$ depends on the potential [see Eq.~(\ref{eq:geupdated})], the differences between the NFW and simulated potentials introduce additional deviations through $g(E)$, thereby causing $\delta_{\log f(E)}$ to exceed $\delta_{\log(dM/dE)}$. As in the case of the energy distributions, for haloes with large $\Phi_{\rm diff}(0)$, the NFW fits show significant deviations from the simulated DFs at $E\lesssim\Phi_\mathrm{NFW}(0.05\rvir)$.

\section{DARKexp Fits to Energy Distribution, DF, and Density Profile}
\label{sec:DARKexp}
In this section, we discuss the DARKexp energy distributions \citep{2010ApJ...722..851H}. We also derive the corresponding DFs and density profiles, and compare them with simulations and the NFW fits. Because the simulated energy distributions are constructed with particles inside $\rvir$, for comparison, the original DARKexp energy distribution should be modified to have a high-energy cutoff. For simplicity, we adopt the following modified form:
\begin{equation}
\frac{dM}{dE}=
\begin{cases}
\begin{aligned}
        A\left[\exp\left(\frac{E-\Phi_0}{\sigma^2}\right)-1\right], && E\leq 0,\\
        \left(\frac{dM}{dE}\right)_{E=0}\left(1-\frac{E}{E_\mathrm{max}}\right),&& E>0,
\end{aligned}
\end{cases}
\label{eq-darkdmde}
\end{equation}
where $(dM/dE)_{E=0}=A[\exp(-\Phi_0/\sigma^2)-1]$ is the value of $dM/dE$ at $E=\Phi(\rvir)=0$. In fitting the above form to the simulated $dM/dE$, we first obtain the best-fit parameters $A$ and $\sigma^2$ by taking the central potential $\Phi_0=\Phi(0)$ from the simulation and minimizing the mean square difference in $\log(dM/dE)$ over the energy range $[\Phi(0.05\rvir),\Phi(\rvir)]$. We then obtain the parameter $E_{\rm max}$ by requiring $\int_{\Phi_0}^{E_{\rm max}}dE(dM/dE)=\mvir$, where $\mvir$ is the virial mass of the simulated halo.

\subsection{Comparison of Fitted and Simulated Energy Distributions, DFs, and Density Profiles}
We compare the best-fit $dM/dE$ in Eq.~(\ref{eq-darkdmde}) with the simulated energy distributions for the eight representative haloes in Fig. \ref{fig:dMdE}.
It can be seen that the DARKexp fits describe the simulated results very well over the energy range $[\Phi_{\mathrm{NFW}}(0.05\rvir),\Phi_{\mathrm{NFW}}(\rvir)]$. Specifically, the average root mean square error in $\log(dM/dE)$ is $\delta_{\log(dM/dE)}=0.08$~dex across our entire selected sample of 79 haloes. The error increases to $\delta_{\log(dM/dE)}=0.12$~dex for the somewhat wider energy range 
$[\Phi_{\mathrm{NFW}}(0.05\rvir),\Phi_{\mathrm{NFW}}(2.5\rvir)]$. These errors are smaller than but comparable to those (0.11 and 0.14~dex, respectively) for the NFW fits.
However, because the DARKexp fits use the simulated central potentials $\Phi_0=\Phi(0)$ as input, they describe the simulated energy distributions at $E<\Phi_{\mathrm{NFW}}(0.05\rvir)$ much better than the NFW fits for haloes with significant $\Phi_{\rm diff}(0)=\Phi_{\rm NFW}(0)-\Phi(0)$.

For the best-fit $dM/dE$ in Eq.~(\ref{eq-darkdmde}), we can determine the corresponding $\rho(r)$ and $f(E)$ from an iterative procedure (e.g., \citealt{10.1093/mnras/200.4.951}) using Eqs.~(\ref{eq:densityfromfe}) [with $\Phi(\infty)$ replaced by $E_{\rm max}$], (\ref{eq:dmde}), and (\ref{eq:geupdated}). We focus on $\rho(r)$ for $r\in[0,\rvir]$, which is sufficient to determine $f(E)$ for $E\in[\Phi_0,E_{\rm max}]$. 
The converged $\rho(r)$ gives the correct central potential $\Phi_0$ and the correct virial mass $\mvir$.
Note that although the $dM/dE$ in Eq.~(\ref{eq-darkdmde}) is fitted to the simulated result using particles inside $\rvir$, it contains information about particles with $E>\Phi(\rvir)$ that can move outside $\rvir$. The corresponding $\rho(r)$ and $f(E)$ are not sensitive to the selection of particles inside $\rvir$, which mainly affects the density of energy states [see Eq.~(\ref{eq:geupdated})].
In other words, had we fitted the DARKexp energy distribution using particles inside e.g., $2.5\rvir$, we would have obtained essentially the same $\rho(r)$ and $f(E)$. 

We compare the DARKexp fits to $\rho(r)$ with the simulated results and the NFW fits in Fig.~\ref{fig:density}.
It can be seen that for $r\in[0.05,1]\rvir$, the DARKexp fits to $\rho(r)$ have the same quality as the NFW fits, with an average root mean square error of $\delta_{\log\rho}=0.07$~dex across our entire selected sample of 79 haloes. However, because the DARKexp fits must reproduce the simulated central potentials, they are closer to the simulated $\rho(r)$ at $r<0.05\rvir$ for those haloes with significant $\Phi_{\rm diff}(0)=\Phi_{\rm NFW}(0)-\Phi_{\rm sim}(0)$. The comparison of $\Phi(r)$ (Fig.~\ref{fig:potential}) is similar to that of $\rho(r)$. Likewise, the comparison of $f(E)$ (Fig.~\ref{fig:fE}) is similar to that of $dM/dE$ (Fig.~\ref{fig:dMdE}). Table~\ref{tab:error} summarizes the average root mean square errors of the NFW and DARKexp fits to $\rho(r)$, $dM/dE$, and $f(E)$ across our entire selected sample of 79 haloes. 
\begin{table}
	\centering
	\caption{Average root mean square errors of the NFW and DARKexp fits to $\rho(r)$, $dM/dE$, and $f(E)$ across the entire selected sample of 79 haloes.}
	\label{tab:error}
    \setlength{\tabcolsep}{3pt} 
    \renewcommand{\arraystretch}{1.2} 
	\begin{tabular}{llcc} 
		\toprule
            Error & Range & NFW & DARKexp \\
        \midrule
            $\delta_{\log\rho}$  &  $[0.05\rvir,\rvir]$&0.07&0.07\\
            $\delta_{\log(dM/dE)}$ & $[\Phi_{\rm NFW}(0.05\rvir),\Phi_{\rm NFW}(\rvir)]$&0.11&0.08\\
            $\delta_{\log(dM/dE)}$ & $[\Phi_{\rm NFW}(0.05\rvir),\Phi_{\rm NFW}(2.5\rvir)]$&0.14&0.12\\
            $\delta_{\log f(E)}$  &  $[\Phi_{\rm NFW}(0.05\rvir),\Phi_{\rm NFW}(\rvir)]$&0.24&0.15\\
            $\delta_{\log f(E)}$  &  $[\Phi_{\rm NFW}(0.05\rvir),\Phi_{\rm NFW}(2.5\rvir)]$&0.24&0.17\\
        \bottomrule
	\end{tabular}
\end{table}

\subsection{Connection between DARKexp and NFW Energy Scales}
\label{sec:scale}
\begin{figure*}
\centering
\includegraphics[width=1\textwidth]{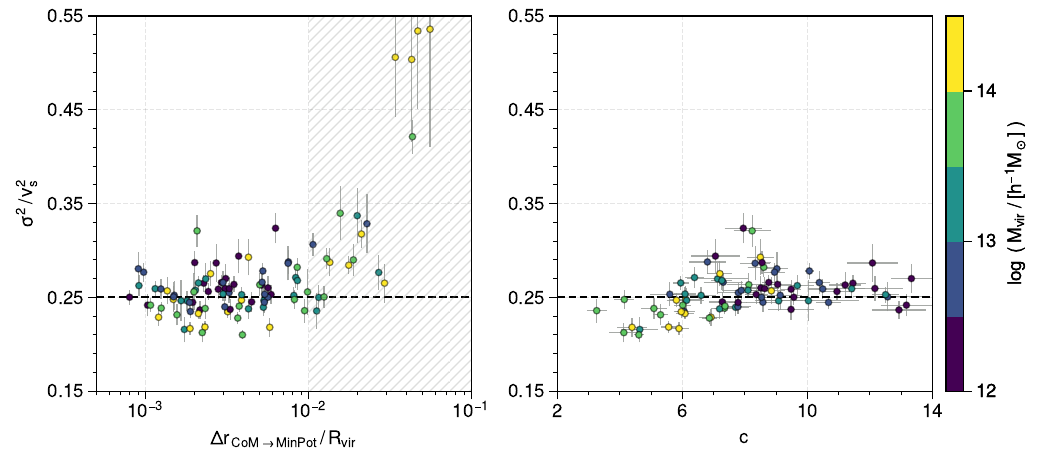}
\vspace{-2em}
\caption{%
\textit{Left panel}: Dependence of $\sigma^2/\vs^2$ on the relaxation level as measured by the offset $\Delta r$ between the center of mass and the location of the minimum potential.
The parameter $\sigma^2$ is the energy scale for the DARKexp fit to the energy distribution, and $\vs^2=4\pi G\rhos\rs^2$ is the energy scale for 
the best-fit NFW profile of the same halo. Each symbol represents a randomly-selected isolated halo.
\textit{Right panel}: The ratio $\sigma^2/\vs^2$ as a function of the concentration $c$ and the virial mass $\mvir$ for isolated and relaxed haloes with $\Delta r<0.01\rvir$.
Error bars are estimated from the fitting procedures.
}
\label{fig:relaxation}
\end{figure*}

The energy scale $\sigma^2$ for the DARKexp energy distribution can be related to the energy scale $\vs^2=4\pi G\rhos\rs^2$ for the NFW profile of the same halo.
As shown in Fig.~\ref{fig:relaxation}, the ratio $\sigma^2/\vs^2$ depends on the relaxation level of the halo.
For offset of $\Delta r<0.01 \rvir$ between the center of mass and the location of the minimum potential, $\sigma^2/\vs^2$ approximately stays around 0.25 and exhibits no clear trend with $\Delta r$, the concentration $c$, or the virial mass $\mvir$.
In contrast, this ratio systematically stays above 0.25 for $\Delta r>0.01 \rvir$ and reaches large values with large errors at large $\Delta r$. 
This behavior suggests that $\sigma^2/\vs^2\sim 0.25$ is established at the end of the relaxation process, and that insufficiently-relaxed haloes have systematically larger $\sigma^2/\vs^2$ with larger errors when NFW and DARKexp fits are used. Therefore, we have only used haloes with $\Delta r<0.01 \rvir$ in the above comparisons of NFW and DARKexp fits with the simulations. 

\section{Discussion and Conclusions}
\label{sec:sum}
Using the best-fit NFW profile of a simulated halo obtained by minimizing the mean square difference in $\log\rho$ for $r\in[0.05,1]\rvir$, 
we have derived the corresponding DF and energy distribution. By matching the NFW and simulated potentials at $\rvir$,
we find that as expected, the NFW fits to the energy distributions and DFs are close to the results for a wide range of haloes in a dark-matter-only simulation (TNG300-1-Dark) from the IllustrisTNG Project (Figs.~\ref{fig:dMdE} and \ref{fig:fE}).
The average root mean square errors are 0.14 and 0.24 dex for energy distributions and DFs, respectively, 
for $E\in[\Phi_{\rm NFW}(0.05\rvir),\Phi_{\rm NFW}(2.5\rvir)]$ across our selected sample of 79 haloes.
Deviations occur at low energy when the NFW profile provides a poor fit for $r<0.05\rvir$. The NFW fits to the DFs have somewhat less accuracy than those to the energy distributions 
because additional deviations are introduced through the density of energy states.

We have also compared the NFW fits to the energy distributions and DFs with the DARKexp fits of \cite{2010ApJ...722..851H} (Figs.~\ref{fig:dMdE} and \ref{fig:fE}). We find that these fits have comparable accuracy
in the region where both fit well (Table~\ref{tab:error}), and that the energy scale for the DARKexp energy distribution is $\sigma^2\sim0.25\vs^2$ (Fig.~\ref{fig:relaxation}), where $\vs^2=4\pi G\rhos\rs^2$ is defined by the parameters $\rhos$ and $\rs$ for the NFW profile. The DARKexp fits are better at low energy because they require the central potential as an input. 
The approximate relation $\sigma^2\sim0.25\vs^2$ (or $\sigma^2\sim \pi G\rhos\rs^2$) may be the end result of the relaxation process, and suggests a relaxation criterion $\Delta r<0.01\rvir$ (Fig.~\ref{fig:relaxation})
for the offset between the center of mass and the location of the minimum potential. This criterion is more stringent than some criteria proposed in the literature (e.g., \citealt{2007MNRAS.378...55M,10.1111/j.1365-2966.2007.12381.x}). While the choice of criterion should depend on the purpose, we propose that $\sigma^2/\vs^2$ can serve as a novel indicator for the relaxation level.

In conclusion, we have studied two approximate methods to obtain energy distributions and DFs of dark matter particles in isolated and relaxed haloes with a wide range of mass.
Based on our results, we suggest that a convenient way of estimating the energy distribution and DF of a halo is to derive them from the NFW profile specified by the virial mass and concentration of the halo.
While the results at $E<\Phi_{\rm NFW}(0.05\rvir)$ may have substantial errors, these errors may not be pertinent in view of the influence of baryonic processes in the central region of the halo.
We have assumed that the energy distribution and DF are functions of energy only, thereby ignoring the anisotropy in the velocity distribution. In the future, we plan to take the velocity
anisotropy into account and derive the corresponding energy distribution and DF from the NFW profile for comparison with simulations.

\section*{Acknowledgments}
This work was supported in part by the US Department of Energy under grant DE-FG02-87ER40328 and by a Grant-in-Aid from the University of Minnesota.
ZL acknowledges the funding from the European Union’s Horizon 2020 research and innovation programme under the Marie Skłodowska-Curie grant agreement No. 101109759 (``CuspCore'') and the Israel Science Foundation Grant ISF 861/20. Some of the figures were created with the colorblind friendly scheme developed by \cite{2021arXiv210702270P}.
\section*{Data Availability}
The data underlying this article will be provided on request to the corresponding author.

\bibliography{draft}
\bibliographystyle{mnras}
\label{lastpage}
\end{document}